\def\ba{\begin{eqnarray}}
\def\ea{\end{eqnarray}}
\def\lb{\label}
\def\nn{\nonumber \\}

\def\g{\gamma}

\documentclass[12pt,nofootinbib]{revtex4-1}
 
\usepackage{graphicx,color}
\usepackage{hyperref}

\begin{document}
\bibliographystyle{apsrev4-1}
\title{Stopped nucleons in configuration space}

\author{Andrzej Bialas}
\email[E-Mail:]{bialas@th.if.uj.edu.pl}
\affiliation{M. Smoluchowski Institute of Physics, Jagellonian University, Lojasiewicza 11, 30-348 Krakow,  Poland}

\author{Adam Bzdak}
\email[E-Mail:]{bzdak@fis.agh.edu.pl}
\affiliation{AGH University of Science and Technology\\
Faculty of Physics and Applied Computer Science\\ 
30-059 Krak\'ow, Poland}

\author{Volker Koch}
\email[E-Mail:]{vkoch@lbl.gov}
\affiliation{
Nuclear Science Division,
Lawrence Berkeley National Laboratory, 1 Cyclotron Road,
Berkeley, CA 94720, USA}

\begin{abstract}
In this note, using the colour string model, we study the configuration space distribution of stopped nucleons in
heavy-ion collisions. We find that the stopped nucleons from the target and the projectile 
end up separated from each other by the distance increasing with the collision energy. In consequence, for the center 
of mass energies larger than 6 or 10 GeV (depending on the details of
the model) it appears that the system created is not in thermal and
chemical equilibrium, and the net baryon density reached is likely not
much higher than that already present in the colliding nuclei. 
\end{abstract}

\maketitle

\section{Introduction}
\label{sec:intro}
Searching of structures in the QCD phase diagram has been at the centre of
the research on the strong interactions for many years. Experimentally,
dense and hot QCD matter is created by colliding heavy ions at ultra
relativistic energies, and experiments at RHIC and the LHC have found
that this matter has remarkable properties, such as a small
shear-viscosity to entropy ratio as well as an unexpected opaqueness
for jets \cite{Adcox:2004mh,Back:2004je,Adams:2005dq}. At the same
time Lattice QCD methods have improved to the point that continuum
extrapolated results for the equation of state
\cite{Borsanyi:2010cj,Bazavov:2014pvz} and also for the nature of the
phase transition at vanishing baryon density \cite{Aoki:2006we} have
been obtained. The
latter was found to be an analytic cross-over transition. Model
calculations on the other hand predict (see
e.g. \cite{Stephanov:2007fk,Gazdzicki:2015ska}) a first order transition at
vanishing temperature and large baryon densities. If this were the
case this first order phase co-existence will have to end at a critical
point, the location of which is not well constrained by neither model
nor lattice calculations. 

The  existence of a critical point and first order phase co-existence
region in the QCD phase diagram has sparked a dedicated experimental
program, the so called RHIC beam energy scan (BES). The basic idea
behind this program is that by lowering the beam energy, one can
create systems at higher average net-baryon density, and indeed experiments
at the CERN SPS and results from the first phase of the BES show
that the observed net baryon number at mid rapidity increases with
decreasing beam energy (see e.g. \cite{Adamczyk:2013dal,Alt:2006dk}). Since no
net-baryon number is being produced in the collision, the only way to
increase the average net baryon density is to transport baryons from
projectile and target to the mid rapidity region. Or in other words
the baryons need to be stopped in the centre of mass frame. However,
stopping the baryons is not enough. In order to have matter at high
baryon density the stopped  baryons {\it from both nuclei} also need to be located within the
 same (and small enough) region  in {\em configuration} space. That this is not so easy to achieve will
be the central point of this note. 

This paper is organised as follows: In the next section we will set
up the formalism based on a simple string picture. Next
we will show some results before we conclude.

\section{Configuration space distribution of stopped nucleons}

As already pointed out in the introduction, studies of the new states of matter created in collisions of heavy ions must involve the discussion of the particle distributions in configuration space. As these distributions cannot be directly measured, one has to rely on indirect methods or on models of particle production.  
Here we study the distribution along the beam ($z$) direction of
nucleons which, after collision of two heavy nuclei,  lost a large
part of their momentum and are located in the mid-rapidity region,
i.e. close to the c.m. rapidity, $y = 0.$  

Denoting by $\sigma$ the energy loss per unit length one has
\ba
dE_z=-\sigma dz
\ea 
where  $E_z$ is the nucleon energy (depending on $z$) and $z$ is its position along the beam axis.
Assuming $\sigma$ to be a constant, as is the case of the Lund model \cite{anderson} and in the bremsstrahlung mechanism \cite{stodolsky} one obtains
\ba
E_z= E_{i}- \sigma (z-z_c)   \lb {lu} 
\ea
where $E_{i}$ is the initial energy and $z_c$ is the collision point in
configuration space. 

In the c.m. frame,  the two nucleons which are to collide, move  right and left with velocity $V$, defined by the initial energy $E_{i} = \frac12 \sqrt{s} = \sqrt{P^2+M^2}$. We denote their positions  at $t=0$ by   
$\zeta_L$ and $\zeta_R$.  

The  collision space-time point $(z_{c},t_{c})$ satisfies 
\ba
z_c=\zeta_R+Vt_c;\;\;\;\;z_c=\zeta_L-Vt_c
\ea
giving
\ba
Vt_c=\frac{\zeta_L-\zeta_R}2;\;\;\;\;z_c=\frac{\zeta_L+\zeta_R}2 .\lb{zlr}
\ea

It is straightforward to relate $z_c$ and $t_c$ to the position of the nucleon $z$, its rapidity $y$, and the time at which it arrives at $z$. From (\ref{lu}) we have 
\ba
z-z_c=\pm\frac{E_{i}-E_z}{\sigma}=\pm\frac{E_{i}-M_\perp\cosh y}{\sigma}  \lb{zcc}
\ea
where the sign depends on the direction of the nucleon (left or right) and $M_\perp^2 = M^2 + P_\perp^2$.
The time when the nucleon arrives at the point $z$ can be obtained using the equation of motion
$dP/dt=-\sigma$ giving 
\ba
t-t_c=\frac{P_{i}-P_z}\sigma= \frac{P_{i}-M_\perp \sinh y}{\sigma} \lb{tcc}
\ea

Now that we have an expression for the space and time distance from the collision space-time 
point for the right and left moving particles in a nucleon-nucleon
collision, we  turn to nucleus-nucleus collisions. To this end we
first derive the distribution of collision points, given the distributions
of nucleons in the colliding nuclei. 

The distributions of the nucleon positions at $t=0$, given by the nuclear shapes, depend on $(\zeta_L-\zeta_{L0})$ and $(\zeta_R-\zeta_{R0})$  
where the subscript $0$ denotes the position of the centre of the
nucleus (at $t=0$).  Using (\ref{zlr}) one obtains the distribution of
collision points
\ba
F_{c}(z_c)&=&\int d\zeta_L d\zeta_R G_L[\g(\zeta_L-\zeta_{L0})]
G_R[\g(\zeta_R-\zeta_{R0})]\delta(\frac{\zeta_L+\zeta_R}2-z_c)\nn 
&=&2\int d\zeta_- G_L[\g(z_c - \zeta_- - \zeta_{L0})]G_R[\g(z_c+\zeta_- -\zeta_{R0})]  \label{coll_dist}
\ea
where $\zeta_\pm=(\zeta_R \pm \zeta _L)/2$ and $\gamma$ is the Lorentz factor. 
Here $G_{L}$ and $G_{R}$ are the distributions 
of the left and right going nuclei, respectively.
It  should be  realised that this formula treats the multiple collisions of a nucleon as independent collection of nucleon-nucleon collisions. Although this is a crude approximation, we feel that it is sufficient for  our semi-quantitative study.  

The distribution of the positions $z$, where the particles are
decelerated to rapidity $y$ for the right-moving, $P_{R}$, and the left-moving nucleons,
$P_{L}$, are then readily obtained from the collision point distribution, Eq.~(\ref{coll_dist}), by using Eqs.~(\ref{zcc}) 

\begin{eqnarray}
P_{R}(z;y)&=&F_{c}\left(z-\frac{E_{i} -M_\perp\cosh y}{\sigma} \right) \label{PR_PL} \\ \nonumber
P_{L}(z;y)&=&F_{c}\left(z+\frac{E_{i} -M_\perp\cosh y}{\sigma} \right)
\end{eqnarray}  
and 
\begin{eqnarray}
P(z;y) = P_{R}(z;y) + P_{L}(z;y)
\label{final_dist}
\end{eqnarray} 
is then the distribution of the space points where the
incident nucleons from the projectile and the target nuclei are decelerated to rapidity $y$ (see the last section for further discussion).

In our semi-quantitative studies we assume Gaussians for $G_{L}$ and $G_{R}$ with central
positions $\zeta_{L0}$, $\zeta_{R0}$ and widths $R_{L}$ and $R_{R}$,  
\ba
G_L(\zeta_L)\sim e^{-(\zeta_L-\zeta_{L0})^2/R_L^2};\;\;\;
G_R(\zeta_R)\sim e^{-(\zeta_R-\zeta_{R0})^2/R_R^2}
\label{gauss}
\ea 

The collision point distribution is consequently given by 
\ba
F_c(z_c)\sim e^{-4\g^2[z_{c}-\Delta_0]^2/(R_L^2+R_R^2)};\;\;\;
\Delta_0=\frac{\zeta_{R0}+\zeta_{L0}}2
\ea
and in the following we will take $\Delta_0 = 0$.

\section {Results}

\begin{figure}[t]
\begin{center}
\includegraphics[scale=0.4]{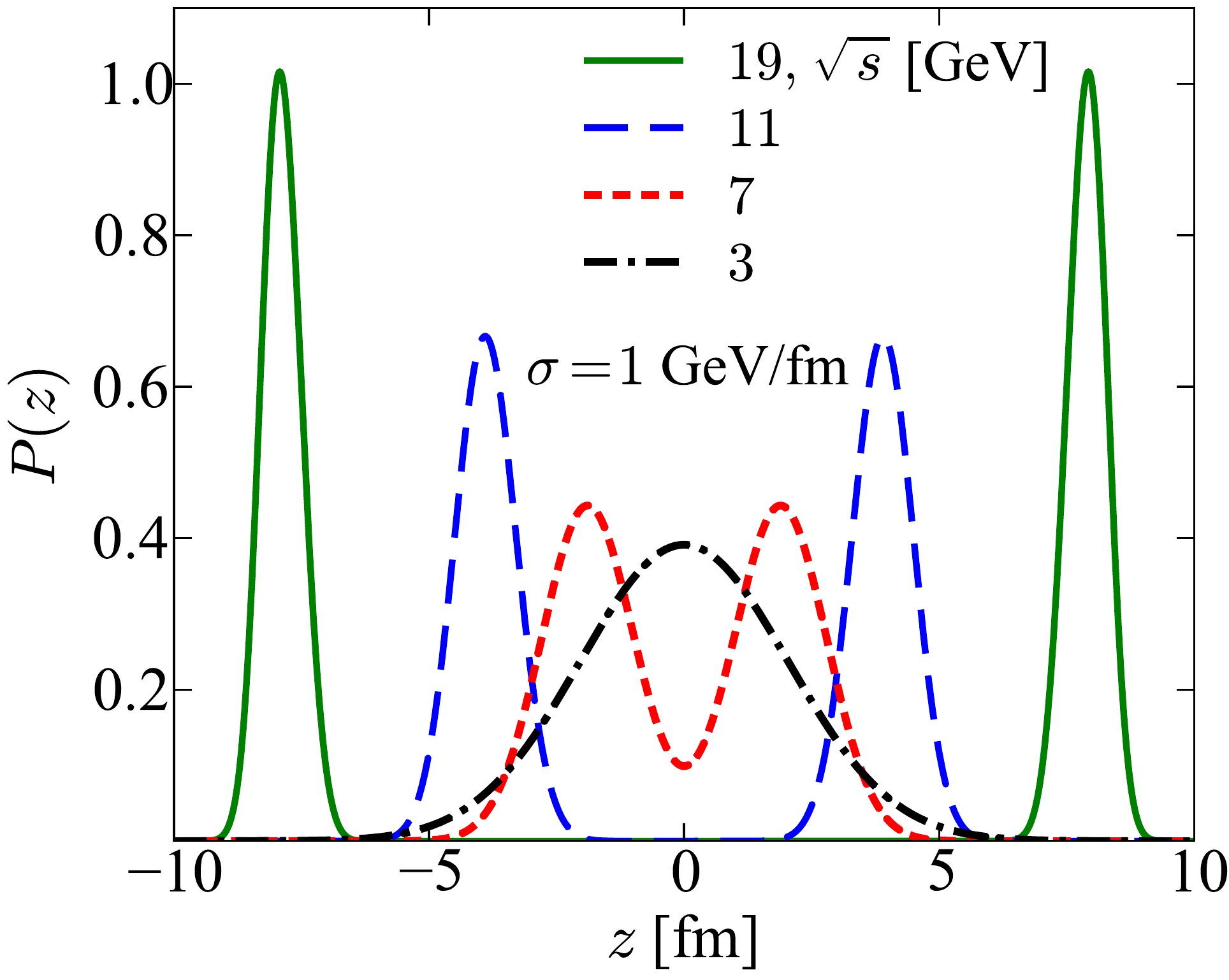}
\end{center}
\par
\vspace{-5mm}
\caption{The $z$ distribution of stopped nucleons at mid-rapidity for $\sigma=1$ GeV/fm (wounded nucleon model) and various c.m. energies, $\sqrt{s} = 3, 7, 11$ and  $19$  GeV/c.}
\label{dz1}
\end{figure}

\begin{figure}[t]
\begin{center}
\includegraphics[scale=0.4]{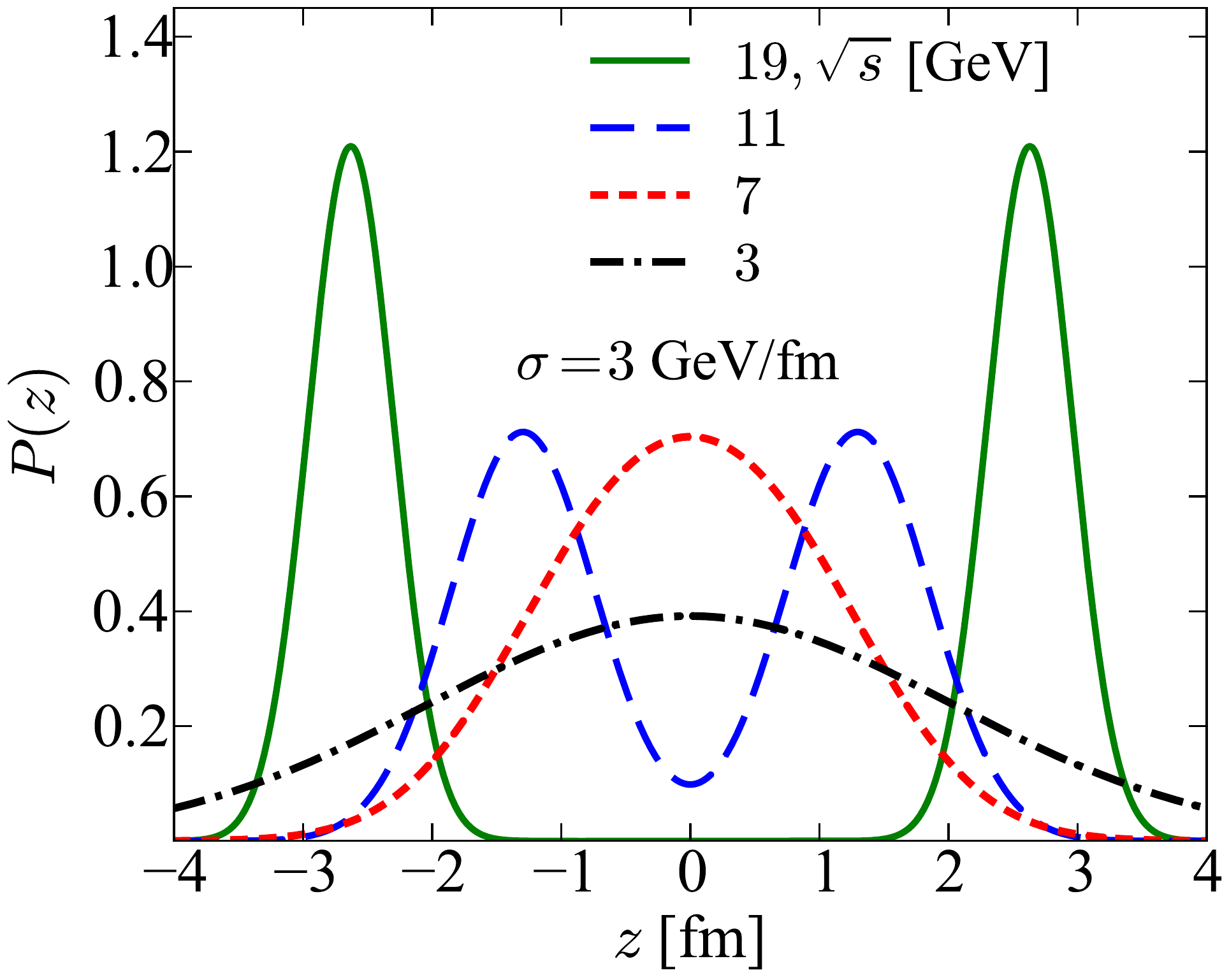}
\end{center}
\par
\vspace{-5mm}
\caption{Same as in Fig. \ref{dz1} but $\sigma=3$ GeV/fm (wounded quark model).}
\label{dz3}
\end{figure}

The distributions of $z$, evaluated for the flat 
rapidity distribution in the region $|y|<1$,\footnote{Our results change weakly when narrowing the rapidity bin.} assuming baryon density per unit of rapidity equal to unity, are shown in Fig. \ref{dz1} ($\sigma=1$ GeV/fm, wounded nucleon model \cite{BBC}) and Fig. \ref{dz3} ($\sigma=3$ GeV/fm, wounded quark model \cite{BCF}). Here $R_L=R_R=6.5$ fm, see Eq.~(\ref{gauss}), and the transverse momentum, $P_\perp$, of the final nucleons was taken 1 GeV/c.  

With increasing energy of the collision the separation between the positions of the left- and the right-movers is increasing while  the width of the two peaks is decreasing. They are clearly separated at energies 
above $\sqrt{s} \simeq $ 10 GeV for the wounded nucleon model and for $\sqrt{s} \simeq $ 20 GeV for the wounded quark model.  

At the same time one sees that the density in the two separated peaks increases rapidly with the incident energy. 
Thus  at these higher energies there are two regions  in configuration space where the baryon density may reach substantial values. 

One should keep in mind, however, that the distributions shown in Figs. \ref{dz1} and \ref{dz3} are normalised to a fixed number (the integrals of $P_R$ and $P_L$ over $z$ equal $1$). Consequently, to obtain  the predicted  baryon density in the configuration space, they must be multiplied by the measured baryon density in the central rapidity region, $|y|\leq 1$. With increasing energy the net baryon density in this region decreases and this may substantially reduce the effect. 

Although the nucleons in the two separated peaks of Figs. \ref{dz1} and \ref{dz3} are close enough in space (and time) to interact,  it remains an open and interesting question if they  can eventually form an equilibrated state at high baryon density. Even if this is not the case, however, they may contribute to the elliptical flow $v_2$ of protons measured  by the STAR Collaboration \cite{starv2}. Indeed, in the transverse space they are apparently correlated with the produced particles, since both groups likely reveal a similar "almond" shape. To translate this transverse space correlation into the transverse momentum distribution is not easily achieved, however, as it demands a certain number of collisions to happen before the nucleons fly apart. This question may perhaps be answered by studying various event generators \cite{Bleicher:1999xi}, \cite{Lin:2004en}.

\section{Concluding remarks and comments}

Based on the colour string fragmentation picture, we have studied
the distribution of ``stopped'' nucleons in the longitudinal configuration space $(z)$. We
find that  there are two regions where one may hope to create a baryonic system with net baryon density higher than nuclear density. 

In the central region, $z\approx 0$, where the stopped nucleons from the two colliding nuclei overlap,  the relatively high baryon density can be achieved at low energies. The actual value of this limiting energy varies from   $\sqrt{s} \sim$ 10 GeV  to $\sqrt{s} \sim$ 6 GeV, depending on details of the model.  

In the regions far off $z=0$, Lorentz contraction of nucleons from one
of the colliding nuclei may, at high enough energy,  also produce
relatively large baryon density, provided the number of baryons
stopped at $y\sim 0$ is large enough. Since this number is rapidly
decreasing with increasing energy, however, it seems unlikely that
these regions play an important role, except in events of very high
multiplicity. 

Clearly a longitudinal configuration for higher energies, as seen in Figs. \ref{dz1} and \ref{dz3}, 
does not resemble a system in 
thermal and chemical equilibrium, where the baryons should
be uniformly distributed in configuration space. Therefore, if our
considerations are correct, it may be difficult to extract information
about the properties of dense matter such as it is determined for
example in Lattice QCD.

\bigskip

Several comments are in order.

(i) Equation (\ref{PR_PL}) describes the distribution of the $z$ points where the
incident nucleons from the projectile and the target nuclei are decelerated to rapidity $y$. What we neglect in our discussion is the fact that nucleons stop loosing energy (and reach their rapidity $y$) at different times, not only because they come from different collision points, but also they final rapidities are different (when we consider particles in some finite rapidity bin). Ideally we should study the time evolution (time snapshots) of all decelerating nucleons, however, this is a rather nontrivial problem to tackle analytically.
 
(ii) Although at high energies the process of particle emission is practically the only effective way to transport the colliding nucleons to the central rapidity region, at  relatively low energies we are considering, other mechanisms may be important and cannot be neglected. The most effective one seems to be the excitation of nucleon resonances.\footnote{E.g. production and decay of a baryon resonance of mass $M^*$ in the process $p+p\rightarrow M^*+M^*$ gives the approximate shift of the baryon rapidity which varies from  $\log (M/2M^*)$  (at threshold)  to  $\log (M/M^*)$ (at high energy, $E\gg M^*$)} This may give substantially larger baryon density than that expected from the Lund model.

(iii) When the nuclei pass through each other, the baryon density may of course reach values which exceed the nuclear density. However, even at the energy as low as $\sqrt{s}= 6$ GeV, the rapidities of the incident nucleons exceed 1, and thus they cannot be  observed  in the  present RHIC experiments.  Moreover, it is difficult to see how these nucleons may form an equilibrated system  without important inelastic interactions.  Investigation of this question (as well as  the one  mentioned above)  using  event generators \cite{Bleicher:1999xi}, 
\cite{Lin:2004en} may help to clear up this problem.  

(iv) It remains an open question if, at high energy, the nucleons from one nucleus, arriving to the small region of $z$ (see, e.g., the curves for $\sqrt{s}=19$ GeV in Figs. \ref{dz1} and \ref{dz3}) may actually form an equilibrated baryon system. This is an interesting problem to study using various event generators.

(v) One may worry that  the relative distance between  the nucleons which, according to the Lund model,  stop at a specific position in the $z$-space, is affected by their interaction through the  nuclear forces before they reach the final destination.  We expect  this effect to be small because the time  needed for the nuclear forces  to operate ($\sim 1$ fm) is longer than the time needed  for nucleons (in their rest frame) to travel from the collision point to the point when they stop to radiate. 

(vi) Recent progress in femtoscopy opens the possibility of experimental verification of the ideas presented 
in this work.\footnote{We thank M. Lisa for for bringing this to our attention.} We feel that such measurements could be of  real help in determining if at low and  medium energies  (which are of main interest in investigation of the strongly interacting matter at high density) the colour string model is indeed the dominant mechanism of bringing baryons at rest in the c.m. frame.
 
(vii) It should be pointed out that our argument does not answer the
question {\it how many} nucleons actually do stop at $y\approx 0$.

(viii) Our model essentially scales p+p collisions to A+A
  collisions by folding in the distributions of nucleons in the
  colliding nuclei. In order to account for the increased rapidity
  shifts in p+A and A+A, see, e.g., \cite{Busza:1983rj}, we use the so called wounded quark model,
  which was shown to reproduce the charged particle distributions in
  p+A and A+A systems, see, e.g., \cite{ab,tan,Bozek:2016kpf,Lacey:2016hqy}. 
  Therefore, we consider the results shown in
  Fig.~\ref{dz3} as the more realistic ones.

In  our view the present study, although admittedly based on a simple model and fairly crude approximations,  calls for a detailed investigation of the phase-space density evolution of baryons in the various event
generators, such as UrQMD \cite{Bleicher:1999xi} or AMPT
\cite{Lin:2004en}. Such simulations  are particularly interesting because they should allow to study  fluctuations of the configuration space density of nucleons on the event-by-event basis.

\bigskip
\bigskip

\vspace{\baselineskip} 
\noindent \textbf{Acknowledgments} \newline
{}\newline 
We thank  Mike Lisa for interesting suggestions and Vladimir Skokov for comments.  
This investigation was supported by the Ministry of Science and Higher Education (MNiSW) 
and by the National Science Centre (Narodowe Centrum Nauki), Grants
Nos. DEC-2013/09/B/ST2/00497 and DEC-2014/15/B/ST2/00175. 
This work was also supported by the Office of Nuclear Physics in the US
Department of Energy's Office of Science under Contract No. DE-AC02-05CH11231.

 \end{document}